\documentclass[11pt,preprint]{aastex}

\begin{document}

\title{The Complex Structure of the Multi-Phase Galactic Wind in a Starburst Merger}
\author{Hsin-Yi Shih, David S. N. Rupke\affil{Institute for Astronomy, University of Hawaii, 2680 Woodlawn Dr, Honolulu, HI 96822  {\it hsshih@ifa.hawaii.edu, drupke@ifa.hawaii.edu}}}

\begin{abstract}

Neutral outflows have been detected in many ultraluminous infrared galaxies (ULIRGs) via the Na I D $\lambda\lambda 5890, 5896$ absorption-line doublet. For the first time, we have mapped and analyzed the 2-D kinematics of a cool neutral outflow in a ULIRG, F10565+2448, using the integral field unit (IFU) on Gemini North to observe the Na I D feature. At the same time we have mapped the ionized outflow with the [\ion{N}{2}] and H$\alpha$ emission lines. We find a systemic rotation curve that is consistent with the rotation of the molecular disk determined from previous CO observations. The absorption lines show evidence of a nuclear outflow with a radial extent of at least 3 kpc, consistent with previous observations. The strength of the Na I D lines have a strong, spatially resolved correlation with reddening, suggesting that dust is present in the outflow. Surprisingly, the outflow velocities of the neutral gas show a strong asymmetry in the form of a major-axis gradient that is opposite in sign to disk rotation. This is inconsistent with entrained material rotating along with the galaxy  or with a tilted minor-axis outflow. We hypothesize that this unusual behavior is due to an asymmetry in the distribution of the ambient gas. We also see evidence of asymmetric ionized outflow in the emission-line velocity map, which appear to be decoupled from the neutral outflow. Our results strengthen the hypothesis that ULIRG outflows differ in morphology from those in more quiescent disk galaxies.

{\it Subject Keywords:} galaxies: individual (F10565+2448) -- galaxies: interactions -- galaxies: kinematics and dynamics
\end{abstract}



\section{Introduction}
Galactic winds, driven by supernova-heated gas \citep{chevalier85} or AGNs \citep{veilleux05}, have been frequently observed in starburst galaxies \citep[e.g.,][]{lehnert95}. They play a significant role in enriching the intergalactic medium with metals and suppressing the star formation and growth of super massive black holes \citep[][and references therein]{veilleux05}. They are an indispensable element in theoretical models of galaxy evolution, in which they provide the feedback mechanism that enables the models to reproduce galaxy scaling relations consistent with those observed \citep[e.g.,][]{dutton09}. Winds have also been invoked to explain a number of issues related to contemporary cosmology \citep[e.g.,][]{veilleux04}, such as the discrepancies between the observed and predicted galaxy luminosity functions \citep[e.g.,][]{springel03}. 

Ultraluminous infrared galaxies (ULIRGs, defined to have $L_{IR} \ge 10^{12} L_{\odot}$) are the late stages of a merger of two gas-rich galaxies, which may then form a QSO (Sanders et al. 1988). Though rare at low redshift, their co-moving density increases rapidly at high redshift \citep{lefloch05}. They are hosts to intense starbursts and may dominate the star formation in the universe at $z \ge 1$ \citep{lefloch04, caputi07}.  Detailed absorption-line studies of a large number of ULIRGs have shown that winds are ubiquitous in these systems \citep{heckman00, rupke02, rupke05a, rupke05b, rupke05c, martin05, martin06} and may play a crucial role in their evolution. 

Our current knowledge of ULIRG winds, especially their geometry, is very limited. Though previous studies of ULIRGs  have yielded valuable insights (including velocity distributions, detection rates, and rough mass estimates), they give little or no information on the spatial variation of wind properties (though see Martin 2006 for a 1-dimensional study). Previous work on estimating properties of ULIRG winds assumed spherical symmetry of the wind with uniform velocity and density \citep{rupke05b}. Since ULIRG winds occur in merging systems that have more complicated dynamics than isolated disk galaxies, interaction between two galaxies can possibly affect the outflow. For example, \citet{rupke05b} found that winds are detected twice as often in ULIRGs as in LIRGs, which suggests that the ULIRG winds may have a wider opening angle compared to winds from less luminous galaxies. If this is true, then a spherically symmetric wind model may be too simplistic to represent ULIRG winds.

To gain a better understanding of ULIRG winds, we obtained 2-D spectra of F10565+2448 with the integral field unit (IFU) on Gemini North. Outflows in our target have been studied with 1-D spectra. \citet{martin06} have studied the velocity variations in this galaxy via long-slit spectra, and \citet{rupke05b} estimated the mass, energy and momentum in the wind. Using the IFU data we can: 1) measure how extended the outflow is, 2) determine whether the basic model for a disk wind \citep[][and references therein]{veilleux05} applies to the late stage of galactic merger, 3) map the velocity structures of different phases of the outflow, and 4) refine estimates of the mass and energetics of the neutral phase, in which most of the mass likely resides.

\section{Observation and Data Reduction}

\begin{figure*}[t]
\centering
\includegraphics[width=5.in]{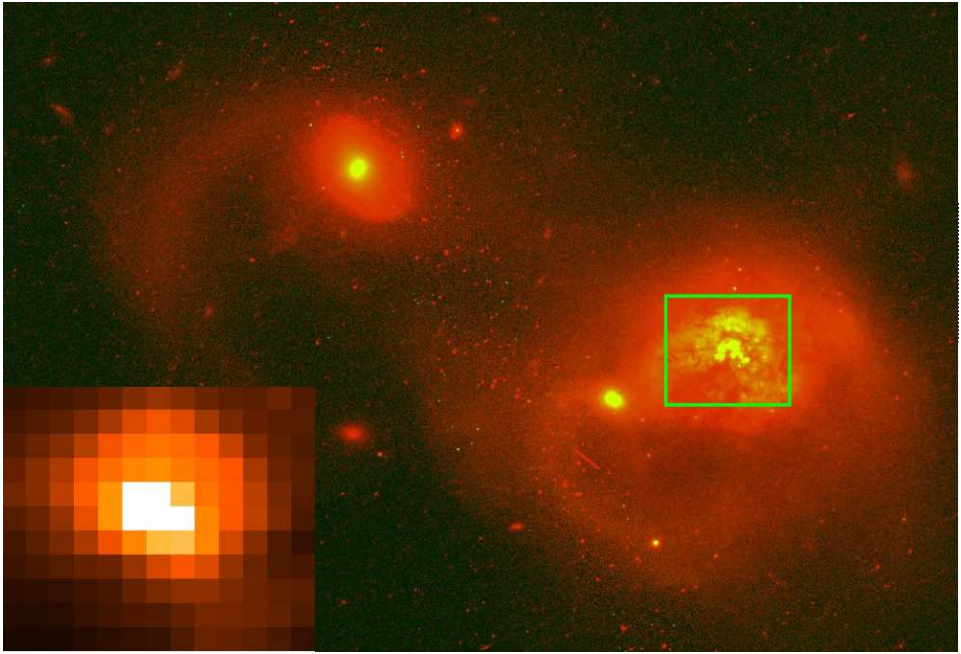}
\caption{HST/ACS two color image showing 3 nuclei and tidal features indicating an ongoing merger.  F435W is shown in green and F814W is shown in red. A green box indicates GMOS field of view and the continuum image of our data is at the lower left. The diagonal stripe going through the middle tis due to noise in the inter-chip region of the detector. The continuum image is summed over wavelength range 6400\AA - 6500\AA. For this and all the following maps, north is up and east is to the left. }
\label{cont}
\end{figure*}

F10565+2448 is one of the nearest and brightest starburst ULIRGs selected from the survey of \citet{rupke05a}, with z = 0.0431 \citep{downes98} and $L_{IR(8-1000\mu\mathrm{m})} = 1.1 \times~10^{12} L_{\odot}$ \citep{sanders03}. The CO redshift measured by \citet{downes98} is consistent with the optical redshift measured by \citet{rupke05a}. It is classified as an \ion{H}{2} galaxy by its spectral type \citep{veilleux95}. However, in a more recent study, \citet{yuan10} classified it as a starburst-AGN composite galaxy, and the AGN contribution to infrared luminosity is estimated to be 10-20\% \citep{veilleux09b, yuan10}. The outflow's absorption-lines' structure in this galaxy shows spatial variation in long-slit spectra \citep{martin06}. An HST image shows that our target has three nuclei, and the one we observed is by far the most luminous. Figure \ref{cont} shows the HST image and a green box indicating the nucleus we observed. 

We observed this object with the Gemini Multi-Object Spectrograph (GMOS) Integral Field Unit (IFU) in two-slit mode using the B600 disperser centered on 6300 \AA, which give us a wavelength range of $5600 - 6980 \AA$ with spectral resolution of 4.26 unbinned pixels (roughly $1.75 \AA$).  Each image covers a science field of $5\farcs ~\times~7\farcs$ , which at this redshift is about $4.2~\times~5.9$ kpc, with 1000 lenslets. There were two dithering positions with offsets of 1\farcs5 from the center in the east-west direction. With eight exposures of 1800 seconds each, a total of four hours worth of data was collected. Seven frames were taken on the night of 2007 Feb 12 (UT) and one was taken on 2007 Jul 4 (UT). The seeing was variable within the range of 1\farcs0 - 1\farcs3. 

The data were reduced using the Gemini IRAF package (version 1.8) following the standard steps: overscan subtraction, bias subtraction, flat-fielding, wavelength calibration, sky subtraction, flux calibration, and spectra extraction. The reduced 2-D images were assembled into a 3-D data cube with coordinates (x, y, $\lambda$) using the GFCUBE routine. For each data cube, differential atmospheric refraction was corrected by shifting the image slices at each wavelength to keep the centroid of the nucleus constant.  After centering and combining all the cubes, and then trimming off the uneven edges, the total science field that was used for analysis is $7\farcs8 \times 6\farcs6$ or $6.5~\times~5.5$ kpc. The original resolution of the images was 0\farcs2 per pixel. To get better signal to noise we binned the data cube to 0\farcs6 per pixel, and since this is smaller than the seeing, we did not lose significant spatial information. The final combined data cube was sliced into 143 1-D spectra that can be used by our absorption-line fitting routine. 

In our final data cube, there appears to be a resolution offset between the two sides of the field of view. After working with the Gemini Helpdesk, we learned that there is a known resolution difference between the two pseudo-slits that is probably flexure dependent. Since the reason for the offset is still unknown there is no standard procedure to fix the problem. After inspecting the reduced data, we determined that we can make a meaningful analysis without any correction. This resolution difference in the wavelength direction can be up to 10 km/s which could possibly affect our spectral fitting, however, we have checked the fitted parameters and found no clear systematic effects between the two pseudo slits. 

\section{Data Analysis}

Methods for analyzing superwinds using spectroscopy utilize both emission and absorption lines. However emission lines can be contaminated by light from star forming regions, and they have an observational degeneracy between infall and outflows. Absorption lines have two advantages in the case of ULIRGs.  First, emission lines in ULIRGs are dominated by rest-frame starburst component \citep[e.g.][]{veilleux95, rupke05b}, while the absorption-line rest-frame component is typically swamped by the outflow component \citep{rupke05b}. This is at least partly due to the different density dependence of emission and absorption-line strengths. Infalls and Outflows sources can also be easily distinguished for absorption lines, since an absorbing cloud has to be between the observer and the light source. In emission lines, on the other hand, blueshifted  gas could be inflowing or outflowing, depending on the (unknown) geometry. \citet{rupke05a} developed software for detailed analysis of the outflows via fitting the Na I D $\lambda \lambda 5890,~5896$\AA~absorption lines. Because of its low ionization energy, Na I D probes the neutral gas, which carries significant amounts of mass and energy and is easily detectable in these galaxies. We used this software to analyze the properties of the wind in F10565+2448. 

We use the fitting routine, NAFIT,  developed by  \citet{rupke05a} to analyze the Na I D feature. This routine fits multiple velocity components to the Na I D absorption lines by varying the covering factor, optical depth, line width (or more specifically the Doppler parameter b = $\sqrt{2} \sigma$ = FWHM/ $ [2 \sqrt{ln2}]$) and central wavelength of each component, and minimizing $\chi^{2}$ using a Levenberg-Marquardt optimizer. 

For each line, assuming a continuum level of unity, the intensity of the absorption line is given by

\begin{equation}
I(\lambda) = 1 - C_{f}(\lambda) + C_{f}(\lambda)e^{-\tau(\lambda)}, 
\label{intensity}
\end{equation}

where $C_{f}$ is the covering factor, and $\tau$ is the optical depth. Assuming a Maxwellian velocity distribution, $\tau$ can be expressed as a Gaussian:

\begin{equation}
\tau(\lambda) = \tau_{c}e^{-(\lambda - \lambda_{c})/(\lambda_{c}b/c)^{2}}, 
\label{tau}
\end{equation}

where and $\tau_{c}$ and $\lambda_{c}$ are the central optical depth and central wavelength in the line and b is the width of the line.  As the Na I D feature is a doublet, and the ratio of the optical depth of the two lines is determined by atomic physics, the central optical depth of the stronger line is always twice that of the weaker line -- i.e., $\tau_{2,c} = 2 \times \tau_{1,c}$. At any particular wavelength, $\tau(\lambda) = \tau_{1}(\lambda) + \tau_{2}(\lambda)$ where $\tau_{i}(\lambda)$ is given by equation \ref{tau} for each line. We also assume an identical covering factor for both lines in a doublet. With this assumption, both optical depth and covering factor can be constrained simultaneously. For a detailed discussion on accounting for blended lines and overlapping absorbers, see \citet{rupke05a}.  

\begin{figure*}[t]
\centering
\includegraphics[width=6.in]{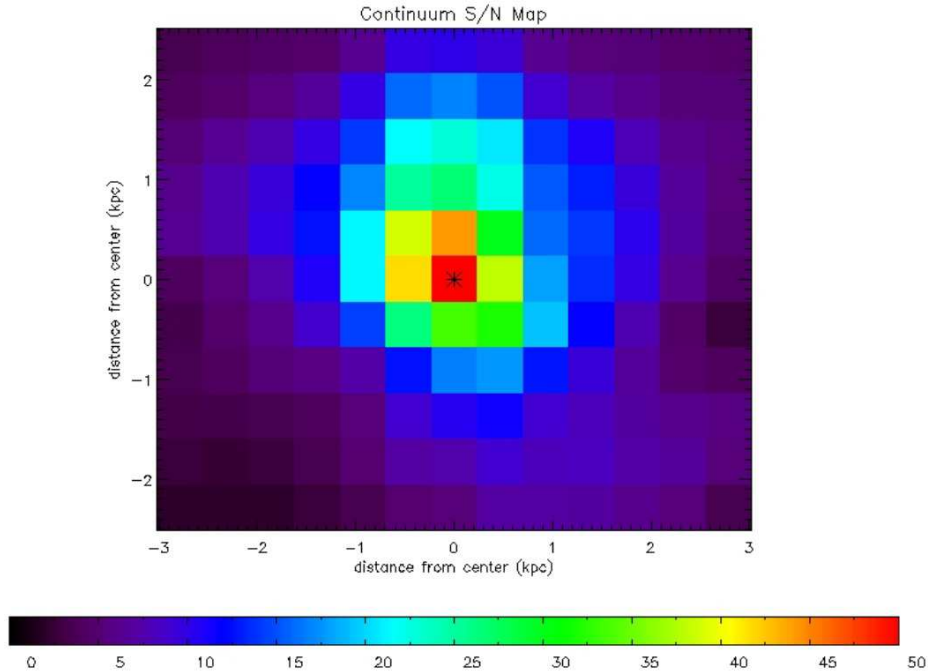}
\caption{A signal to noise ratio map for each spectrum in our data, in units of S/N per unit spectral resolution element. A star marks the continuum peak which we will use as the central reference pixel when referring to any positions in our field of view.}
\label{snimg}
\end{figure*}

\begin{figure*}[t]
\centering
\includegraphics[width=6.in]{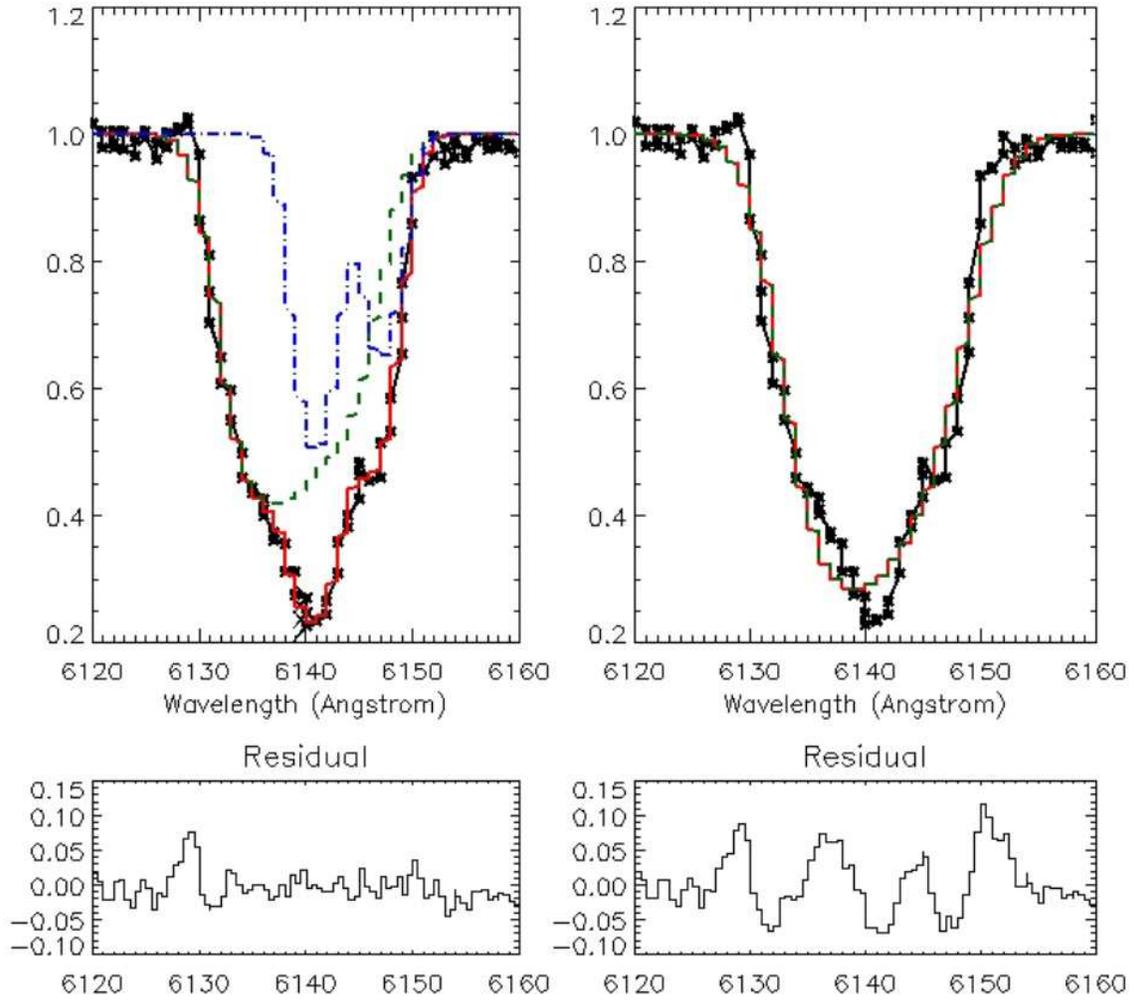}
\caption{A two component fit to the Na I D doublet (left panel) at the continuum peak, a one component fit (right panel), and the residuals. For the two-component fit, the two velocity components are the blue (dash dot) and green (dashed) lines, and the total fit is shown in red (solid). The two component fit clearly leaves less residuals than a one component fit.}
\label{12comp}
\end{figure*}

\begin{figure*}[t]
\centering
\includegraphics[width=6.in]{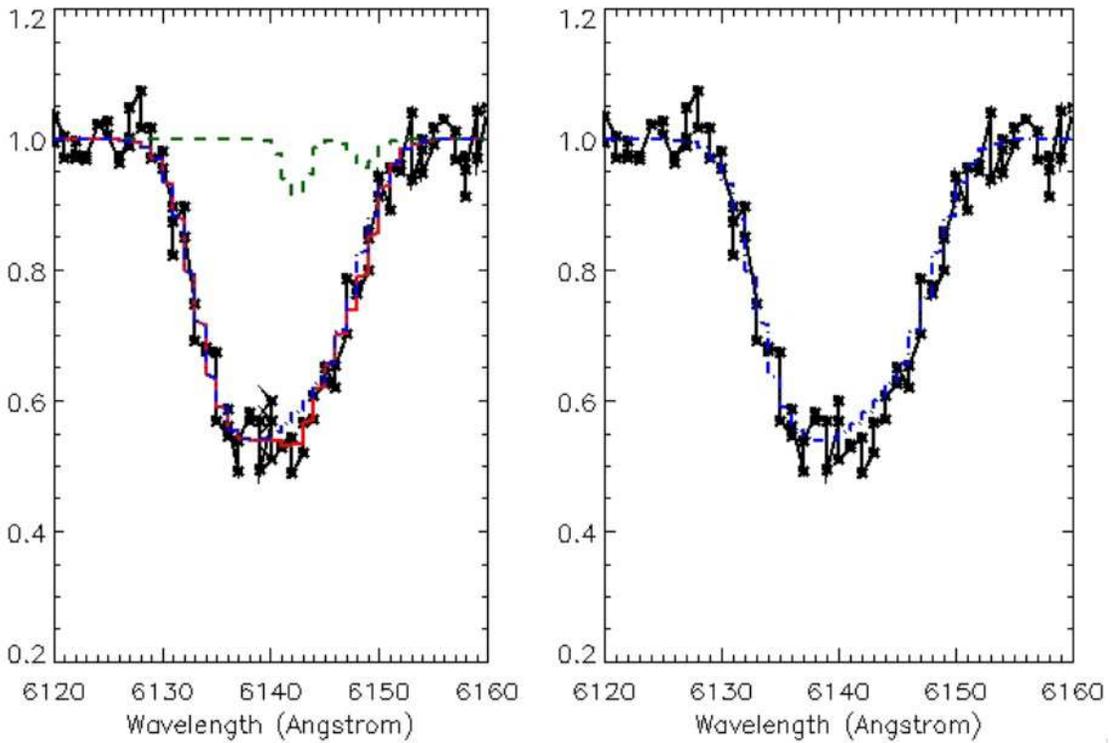}
\caption{A fit that contains only one significant component. The blue (dash dot) and green (dashed) lines (left panel) are the two components and the red line is the total fit. The right panel shows just the data and the blue component which fits the data just as well without the green component. In this case the green component is removed from the final maps. This particular spectrum is taken $\sim 1.8$ kpc from the center.}
\label{sigfit}
\end{figure*}

We wrote a wrapper to this fitting software to fit all of our spectra automatically. Starting at the center of the galaxy, where signal-to-noise is the highest, we step outward in a spiral pattern, using the output parameters of the previous fit as the initial guess for the next spectrum. Fig \ref{snimg} is a signal-to-noise map of the continuum near the Na I D line, in units of S/N per unit spectral resolution element. The S/N is as high as 50 at the center of the nucleus and declines to as low as 2 or 3 at the edges of the FOV. As shown in Fig \ref{12comp}, at the continuum peak, the absorption line is best fitted by two velocity components. (Note: Each velocity component contains a pair of doublet lines). The one component fit can reproduce the basic shape of the line but leaves significant residuals. A two component fit lowered the residuals significantly, and three or more components over-constrains the fit. As we move outward to regions with lower signal-to-noise, we find spectra that can be fit by one component equally well as by two. However, switching between one and two component fits within the same data cube introduces artificial patterns to our velocity maps. Therefore, we first perform two-component fits for all spectra; we then visually inspect the fits and remove any components that have negligible contribution to the equivalent width of the line. Examples of significant vs. insignificant components are shown in Fig \ref{sigfit}.

\begin{figure*}[t]
\centering
\includegraphics[width=6.in]{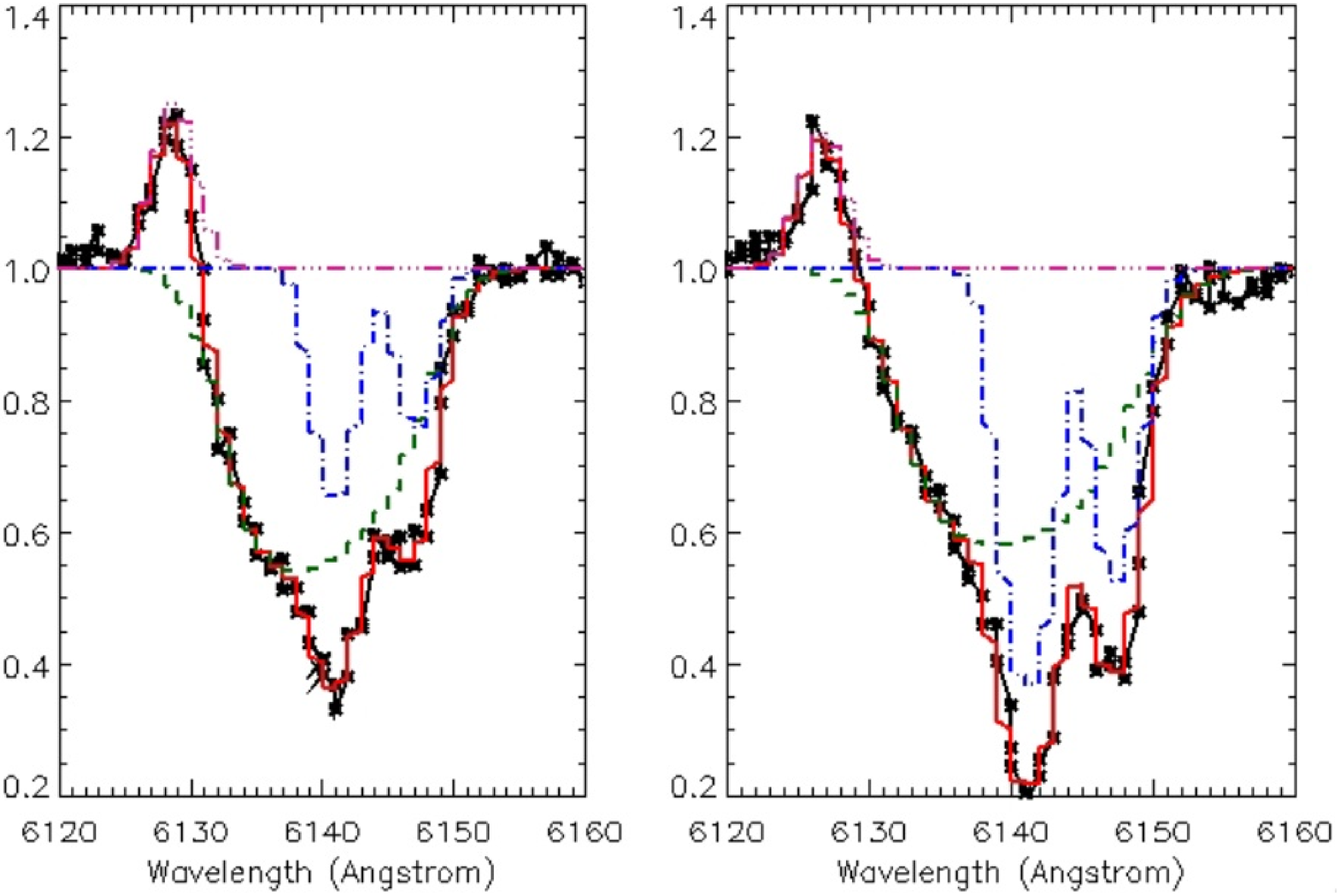}
\caption{Two examples of spectra which require He I 5876\AA~emission-line fits. The pink (dash dot dot) component fits the emission line, reducing its effect on absorption-line fitting. Both spectra are $ \sim 0.5$ kpc from the center. }
\label{hefit}
\end{figure*}

Some spectra show the He I  $\lambda 5876$ emission line which appears immediately blueward of the Na I absorption lines. We include this emission line in our fit, where necessary, to remove its effect on our absorption-line fitting. For most of the spectra, however, the He I emission is weak enough that it blends in with the noise and can be safely ignored. Figure \ref{hefit} shows examples of He I line fitting.

All parameters are allowed to vary freely during the fit. Upper and lower limits of all parameters are the same for all spectra except for the central wavelength which is $\pm~5$ \AA~of the previous fit. The upper limit of the optical depth $\tau$ is set to 5 since our fits are insensitive to optical depths beyond that. The covering factor is constrained to be 0 to 1. To ensure that we are not extracting velocity dispersion widths smaller than the spectral resolution, the lower limit of the Doppler parameter is set to the spectroscopic resolution of $75$ km/s determined from the line width of the calibration lamp lines. The fitted velocity widths are typically well above this lower limit. It should also be noted that \citet{rupke05a} have shown that not de-convolving the instrumental resolution before the fitting does not change the result significantly. The upper limit of the velocity width is set to $500$ km/s and no spectra came close to exceeding this limit.

We have checked our Monte Carlo simulation outputs for possibilities of fitting degeneracies. Since this step essentially fits slightly modified versions of the same spectrum 1000 times, degeneracies for any parameters should show up as a bimodal or more complicated distribution in the fitting results. We find that almost all of the fitted parameters show a roughly Gaussian distribution. Exceptions occur mostly for the lower S/N spectra around the edges that are later discarded anyway due to their low quality. 

We fit 5 emission lines ([OI] 6300\AA~ and 6364\AA, H$\alpha$, and  [\ion{N}{2}] 6548\AA~ and 6583\AA) using a suite of routines that simultaneously estimates the stellar continuum and fits all emission lines at once, with the widths and redshifts constrained to be the same for each emission line.  The stellar continuum, minus emission lines, was estimated using the stellar synthesis models of \citet{gonzalez05}.  The emission lines were fit with two Gaussians, one narrow and one broad; the broad component was discarded near the edges of the data, where the signal-to-noise was too low. The emission-line fitting routines are described further in Rupke et al. (2010, in prep.).

\section{Results}

Figure \ref{cont} is the HST/ACS two color image of our target, with F814W shown in red and F435W shown in green. The green box on the lower right nucleus indicates the GMOS field of view, and the continuum image of our data is shown in the lower left. The continuum flux map of our target is integrated over 6400 - 6500 \AA. As mentioned in Section 2, the pixels are binned $3\times3$ to get better signal to noise.

\begin{figure*}[t]
\centering
\includegraphics[width=5.in]{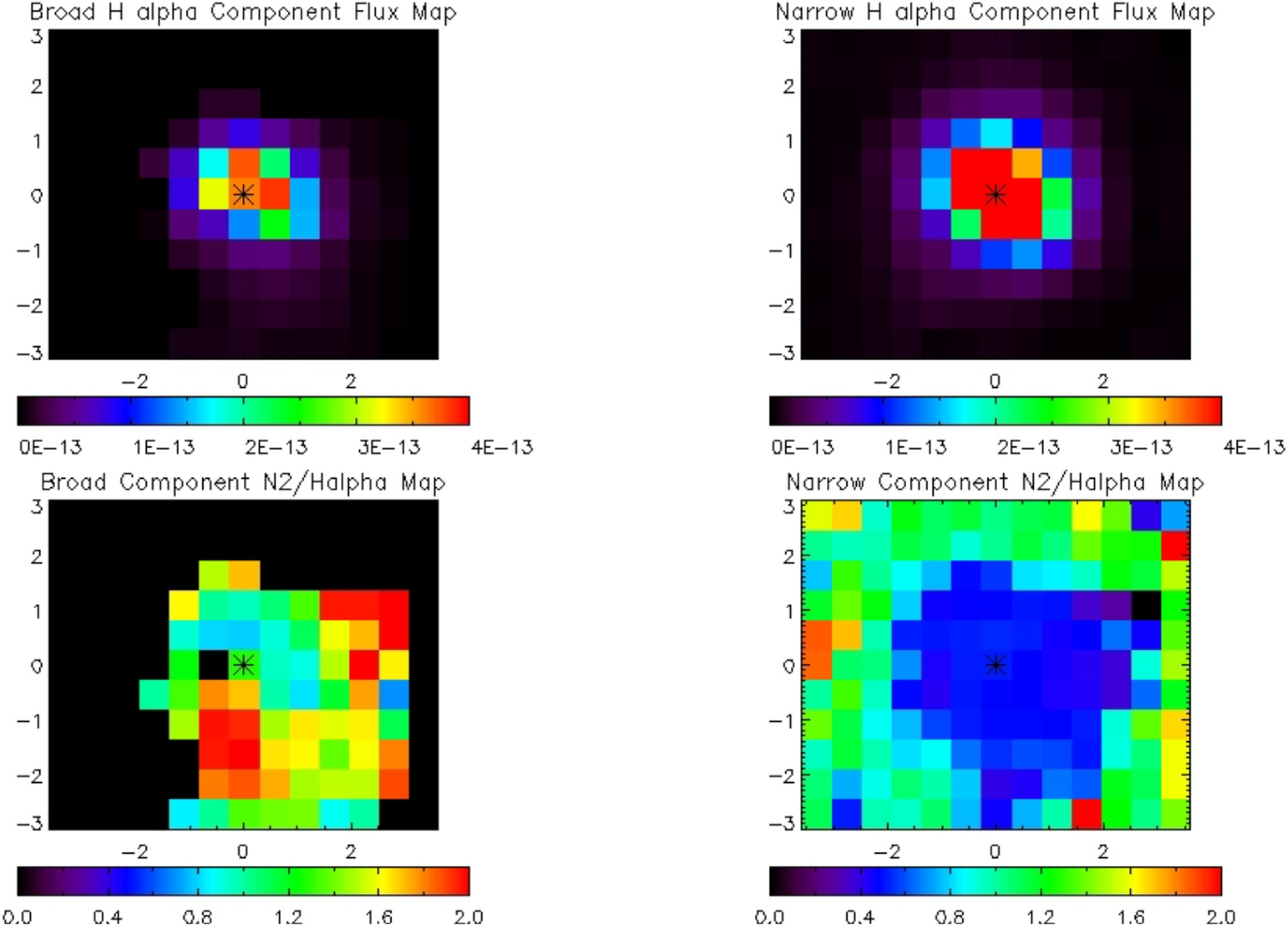}
\caption{Top two panels: H$\alpha$ flux map of both broad and narrow components in units of erg~s$^{-1}$~cm$^{-2}$~arcsec$^{-2}$. Bottom two panels: [\ion{N}{2}]/H$\alpha$ map for both emission components. The narrow component have lower line ratios in the middle and increase outward. The broad component has high [\ion{N}{2}]/H$\alpha$ ratio throughout and a peak in the southeast, possibly due to shock induced ionization. The peak in [\ion{N}{2}]/H$\alpha$ ratio suggests high local density that may impede the outflowing gas. The star marks the pixel with the highest continuum flux. The x and y axes are distance from the center in units of kpc.}
\label{em_lines}
\end{figure*}

The H$\alpha$ flux map is shown in the top two panels of Figure \ref{em_lines}. We've applied a global Balmer-decrement-determined extinction correction to the fluxes, E($B\!-\!V$) = 1.75 \citep{veilleux95}, which corresponds to $A_{V} = 1.75 \times 3.1 = 5.43$. For both components, the H$\alpha$ flux peaks at the same pixel as the continuum flux map. The bottom two panels show the [\ion{N}{2}]/H$\alpha$ flux ratio map for both components. For the narrow component, the line ratio is fairly uniform and symmetric around the center and has an average value of ~0.5, consistent with star formation \citep{kewley06}. Away from the nucleus, however, the ratio increases. The broad component has a much wider range of line ratios and a peak of ~1.8 appears in the south and northwest.

Using the H$\alpha$ flux from the emission line fitting we can estimate a star formation rate (SFR). Integrating over the H$\alpha$ flux map we obtained a total flux of F(H$\alpha$) = 4.08 $\times~10^{-12}$ erg~s$^{-1}$cm$^{-2}$. Using Ned Wright's Javascript Cosmology Calculator \citep{wright06}, assuming $H_{o} = 71$ km s$^{-1}$Mpc$^{-1}$, $\Omega_{M}  = 0.3$ and $\Omega_{vac} = 0.7$, we get a luminosity distance of $D_{L} = 188$ Mpc, and therefore L(H$\alpha$) = 1.7 $\times~10^{43}$ erg/s. SFR can be estimated using the following equations \citep{kennicutt98}:

\begin{equation}
SFR (M_{\odot}~ yr^{-1}) = 7.9 \times 10^{-42} L(H\alpha) (erg~s^{-1})
\end{equation}

\begin{equation}
SFR (M_{\odot}~ yr^{-1}) = 4.5 \times 10^{-44} L_{IR(8-1000\mu\mathrm{m})} (erg~s^{-1})
\end{equation}

The SFR estimated from L(H$\alpha$) is about 136 $M_{\odot}$ yr$^{-1}$, and 190 $M_{\odot}$ yr$^{-1}$ from $L_{IR}$. The SFR estimated from H$\alpha$ flux is slightly smaller as expected since the extinction from \citet{veilleux95} is a lower limit. 

\begin{figure*}[b]
\centering
\includegraphics[width=5in]{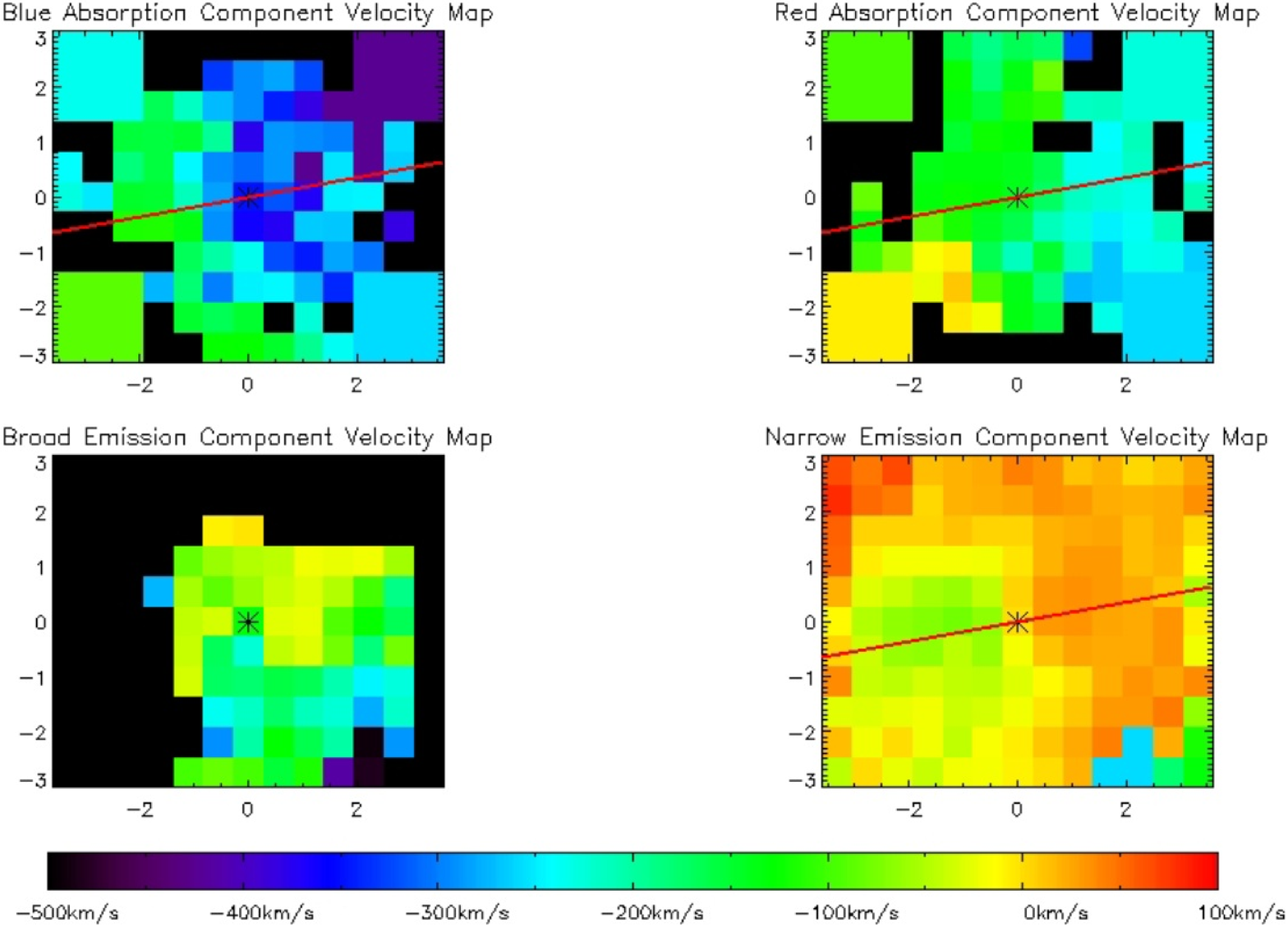}
\caption{Velocities of emission and absorption lines with respect to fixed redshift of $z=0.0431$. The narrow emission line shows rotation consistent with rotation of the molecular disk. The major axis, with PA of 100$^{o}$ east of north \citep{downes98}, is plotted on the top two and bottom right panel. The broad emission-line component is blueshifted south. The absorption lines are blueshifted with east-west oriented velocity gradients opposite to that of the systemic, which is most likely a result of non-uniform density of ambient gas. The four large pixels are the four corners are results of an additional $3 \times 3$ binning done to constrain the fitted parameters at the edges with low S/N.}
\label{velmaps}
\end{figure*}

Figure \ref{velmaps} shows the velocity map of the Na absorption lines and emission lines respectively. (Note: Since the signal to noise is very low at the edges of the FOV, we did an additional $3 \times 3$ binning at the four corners so we can constrain the parameters better. This only applies to the absorption-line maps and not the emission-line maps. This additional binning shows up as four large pixels, which are 9 times larger than the rest of the pixels, at the four corners of all the absorption-line velocity and velocity dispersion maps.) Velocities are calculated with respect to a fixed redshfit of $z=0.0431$. The narrow emission line map, which we interpret as systemic, is blueshifted in the east and redshifted in the west and clearly shows rotation (see Section 5 for more discussion). The PA of the major axis according to \citet{downes98}, 100$^{o}$ east of north, is plotted as red lines on the absorption-lines and narrow emission-line velocity maps as a reference. Most of the northern half of the broad line emission map has values very close to systemic, but the southern half shows significant blue shifts. The absorption lines are separated into higher-outflow-velocity (blue) and lower-velocity (red) components. Both components are generally blue shifted with respect to systemic and their velocity gradients are east-west oriented, roughly aligned with the major axis. 

\begin{figure*}[h]
\centering
\includegraphics[width=5.in]{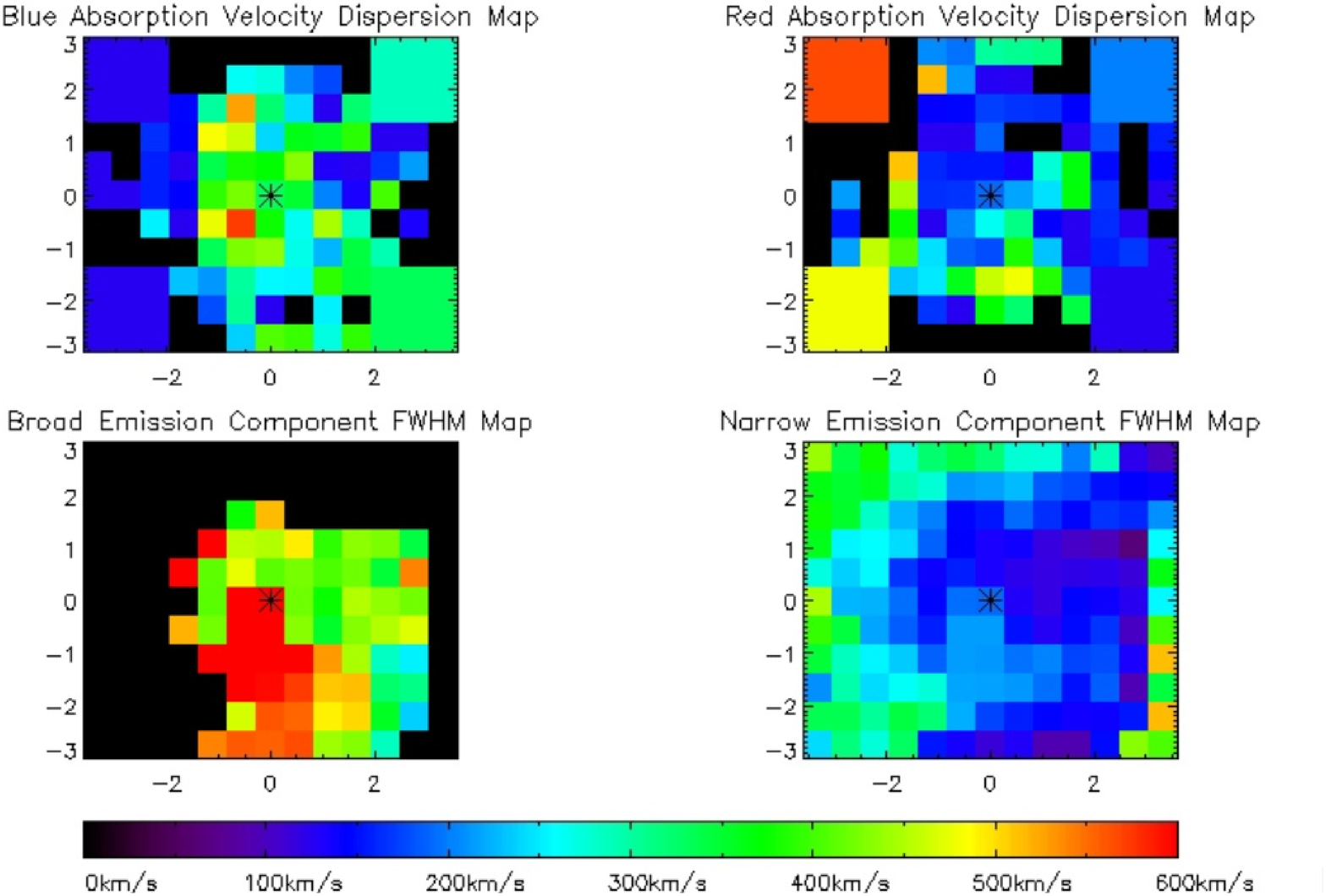}
\caption{Absorption and emission-line velocity dispersion (FWHM) map. For both emission-line components, a high velocity dispersion region appears in the southeast for both components corresponding to the shock-excited region, indicating turbulence caused by wind shocking and mixing with interstellar gas. For the absorption-line components, the higher velocity (blue) component generally has higher velocity dispersion than lower velocity (red) component}
\label{vdisp}
\end{figure*}

Figure \ref{vdisp} shows the velocity dispersion, or full width at half maximum (FWHM), maps of the emission and absorption lines respectively. Both narrow and broad emission lines have a high velocity dispersion region in the southeast, roughly in the same region as the peak of [\ion{N}{2}]/H$\alpha$. The red absorption component has a typical FWHM of roughly 200 km~s$^{-1}$, while the bluer component has a higher typical FWHM at about 300 km~s$^{-1}$. The systemic emission-line FWHM increases away from the nucleus from about 150 to 350 km~s$^{-1}$.

NAFIT calculates the error in each fitted parameter using Monte Carlo simulations described in detail in \citet{rupke05a}. Errors in the velocities are 5-10 km s$^{-1}$ close to the center and remain $\le$ 30 km~s$^{-1}$ for most of the pixels, although for some pixels close to the edge the error can be $>$ 50 km~s$^{-1}$. FWHM errors start from $\le$ 10 km~s$^{-1}$ in the center, remain below 50 km~s$^{-1}$ for most pixels, and increase up to $>$ 80 km~s$^{-1}$ around the edges. 

Other than the random errors due to noise, the systematic effect between the two pseudo slits mentioned in \S\ 2 may also cause errors. The two pseudo slits are oriented north-south with respect to each other. Judging from the images of the fitted parameters, we can see that there is no clear systematic offset in the north-south direction (the direction of the systematic offset). The velocity gradient is oriented in the east-west direction, and from the arc lamp measurements we find that the resolution offset is small (about 10 km~s$^{-1}$) compared to the velocity gradient observed. Therefore we can safely conclude that any errors caused by the systematic offset have a negligible effect on our analysis. 

\begin{figure*}[h]
\centering
\includegraphics[width=5.in]{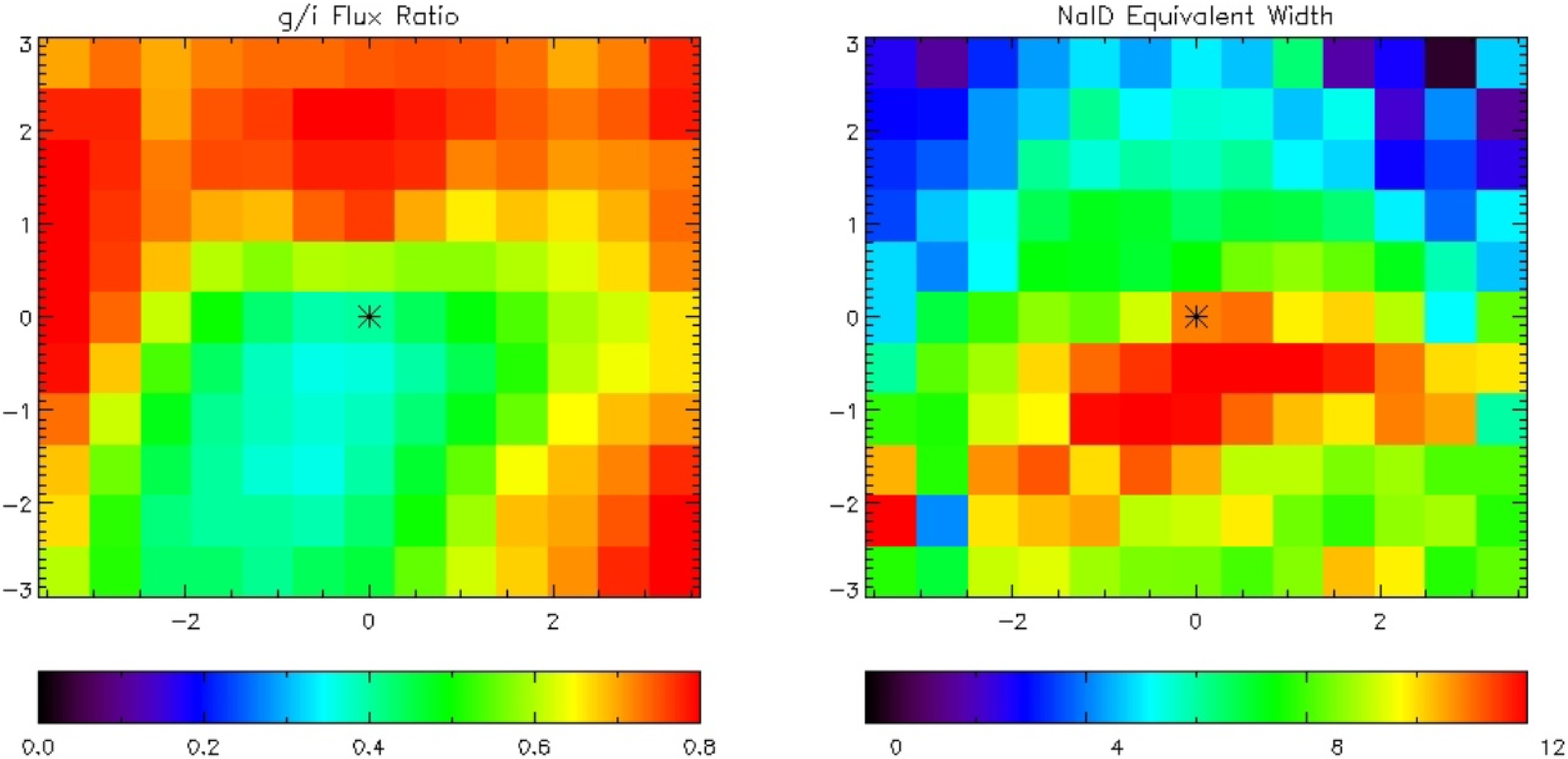}
\caption{Left: SDSS $g/i$ ratio map. SDSS image rotated and binned to match the PA and resolution of our target. Lower $g/i$ ratio indicates the redder and therefore dustier region. Right: Na I D equivalent width map in Angstroms. Higher equivalent width region corresponds to lower $g/i$ flux ratio, implying that dust is present in the entrained outflowing material.}
\label{color_dust}
\end{figure*}

\begin{figure*}[h]
\centering
\includegraphics[width=5.in]{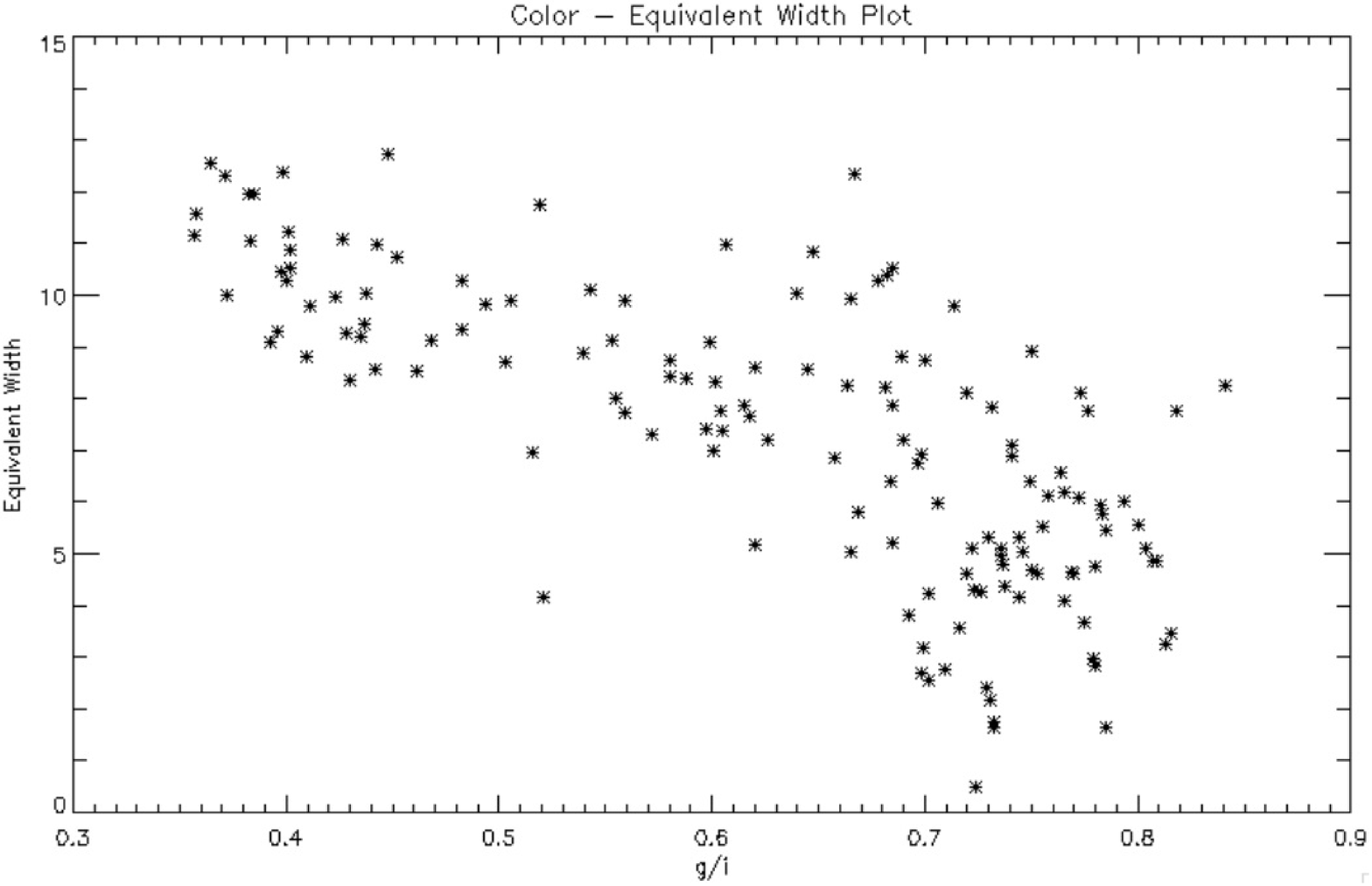}
\caption{A pixel to pixel plot of the SDSS $g/i$ ratio and the Na I D equivalent width, showing spatial correlation between dust reddening and strength of Na I D absorption lines.}
\label{gi_eqw}
\end{figure*}

Our target is known to be dusty and we can test whether the neutral gas absorption corresponds to the dust features. A map of the Na I D line equivalent width is shown in the right panel of Fig \ref{color_dust}.  The equivalent width shown here is the total equivalent width of both velocity components. A map of the dust features is shown in the SDSS $g/i$ ratio map in left panel of Fig \ref{color_dust}. We are using $g/i$ ratio as a proxy for E($B\!-\!V$) and assuming that the stellar population does not vary too much across the field of view. A lower $g/i$ ratio indicates a redder color and corresponds to the easily visible dust filaments in the high resolution HST image (Fig \ref{cont}). The SDSS $g$ and $i$ images are rotated and resampled to match the PA and resolution of our data. A pixel to pixel comparison plot of $g/i$ ratio and equivalent width images is shown in Fig  \ref{gi_eqw}. The Na I D line strength correlates well with the dust reddening. 

We compared our IFU data with the freshly analyzed long-slit observations of our target along a position angle of 53 degrees \citep{rupke05a}. The long slit data probe further from the nucleus, and show the presence of a neutral (ionized) wind up to $7-8$ (10) kpc from the nucleus, as well as regions of large line broadening and shock excitation that extend 10 kpc from the nucleus.  This is consistent with similar results from \citet{martin06} along a PA of 109 degrees. Quantities calculated from \citet{rupke05a} long-slit data, including outflow velocities, velocity dispersions, H$\alpha$ fluxes, and [\ion{N}{2}]/H$\alpha$ ratios, agree with our measurements from the IFU data.

\section{Discussion}
\subsection{The outflow velocity structure}
The narrow line emission shows a clear rotation signature with blue shifted lines in the east and redshifted lines in the west, which is consistent with long-slit H$\alpha$ observations by \citet{martin06} and \citet{rupke05a}. Within errors, this is also consistent with the CO presented in \citet{downes98}. Our velocities ranged from -122 to 35 km s$^{-1}$, while \citet{downes98} measured -70 to 70 km s$^{-1}$, relative to central velocity, in the molecular disk. In general, the narrow emission-line velocities follow the rotation of the molecular disk closely. The small velocity shift across the nucleus is consistent with the galaxy being almost face on in our line of sight (Downes \& Solomon 1998 estimated $i$ = 20 degrees). 

Based on the velocity map, the outflow extends at least 6 kpc in diameter across the center of the galaxy, confirming previous results \citep{martin06}. According to the H$\alpha$ and CO flux distributions, the launching region of the wind is roughly the inner 1 kpc of the galaxy where the starburst and the molecular gas are concentrated. Since the H$\alpha$ flux peak is the same as the continuum peak, we define the launching region as the area within 1 kpc of the continuum peak. Thus, the wind extends at least 2 kpc away from the source region in all directions; this distance is a lower limit if we assume extensions along the line-of-sight.

We see evidence of asymmetric outflow in the broad emission-line and absorption-line velocity maps as mentioned in the previous section. In principle, red shifted emission lines can be observed from wind receding on the opposite side, but they are not observed most likely due to dust obscuration, which is consistent with other ULIRGs observations \citep{veilleux99, rupke05b}. The general pattern of the broad emission-line velocity clearly differs from that of the absorption lines. The broad emission lines only appear to be blue shifted in the southern region while the absorption lines are blue shifted all across the map. The warm ionized gas that produces the blue shifted emission lines we observed arises at boundaries between the wind fluid and ambient gas. Though the relationship between velocities of absorbing-line clouds and line-emitting gas is not well known, previous studies \citep[e.g.,][]{heckman00, rupke05c} have suggested that they may have different velocity structures, which is consistent with our observations. 

The surprising result from the absorption-line velocity maps is that the velocity gradient is opposite to that of the systemic rotation and aligned with the major axis. In general, the narrow emission-line velocity map is blue shifted in the east and red shifted in the west. The outflow velocity maps are less blue shifted in the east and more blue shifted in the west. 

\citet{martin06} observed our target with a long slit going through the continuum peak of the nucleus we observed, and a PA of $109$ degrees. Their data suggested that the velocity of the neutral outflow is roughly constant across the entire galaxy, which implies an outflow diverging from a central starburst and slowing down with increasing height to conserve angular momentum. However, in our observation we do see a significant velocity jump in the absorption lines. Outflows with significant velocity shear along the major axis may be at a low scale height and therefore retain the rotation velocity of the host galaxy, but in this scenario the velocity gradient should be in the same direction as the systemic gradient while the velocity gradient we observed is in the opposite direction.  

One can compare the data to a simple toy model. Consider a galactic wind which consists of a gas disk and central starburst from which the wind is launched, and an outflow that extends perpendicular to the disk and away from the center in a conical shape (see, e.g., Martin 2006 for graphic illustrations). According to this model, an outflow that conserves angular momentum shows very little velocity gradient along the major axis, but may show a velocity jump along the minor axis if the galaxy is tilted with respect to the observer. We do not see a significant velocity gradient in the minor axis direction. However, since the central disk is almost face on, a velocity gradient along the minor axis due to projection effects may not be detectable in our case. 

In the context of this simple model, one could observe a major axis gradient if the entrained outflowing material maintained the rotation it had prior to entrainment.  However, this is inconsistent with the fact that the observed gradient is opposite to disk rotation. We thus conclude that the simple model does not adequately explain our observations.  A wind perpendicular to the central disk is still a possibility, but another ingredient is necessary to explain the major axis velocity gradient.

Another possible explanation of the observed outflow velocity is an uneven power source. If there is a stronger starburst in the west, and this is accelerating the wind at a faster rate, the wind may appear to be counter rotating with respect to the galaxy. But this explanation is inconsistent with the distribution of star formation (H$\alpha$ flux) and the gas reservoir (CO flux), which are both centrally concentrated.   

Our best explanation is that the velocity difference is caused by interaction between the wind and ambient gas. The broad emission component, with high velocity dispersion and [\ion{N}{2}]/H$\alpha$ ratio, is detected up to 2.5 kpc from the nucleus, indicating a widespread shocking of interstellar gas. \citet{west09} found a broad emission-line component in M82 outflow with increasing line width away from the nucleus; they attribute this to turbulence as the wind shocks and mixes with the surrounding gas. If our broad emission-line component is of the same nature, varying densities in the ambient gas may be the cause of velocity asymmetry.  

More evidence that indicates shocks caused by winds can be seen in the broad emission-line [\ion{N}{2}]/H$\alpha$ ratios which ranges from $\sim$0.8 to $\sim$2.0. Such high ratios can only be explained by high velocity shocks \citep{dopita95} or an AGN. Since starburst dominates the luminosity in F10565+2448 \citep{veilleux09, yuan10}, high velocity shocks are the most plausible explanation. Previous work has used the correlation between ionization and velocity dispersion as evidence of shock induced ionization \citep[e.g.][]{veilleux95, dopita95}. Our [\ion{N}{2}]/H$\alpha$ and velocity dispersion maps show that higher ionization corresponds to higher velocity dispersion (bottom left panels of Fig \ref{em_lines}, \ref{vdisp}), and supports a scenario in which outflowing gas interacts with and shocks the surrounding materials. 

Variation of density in surrounding gas can be seen in a localized high ionization region found in the southeast of the broad emission line [\ion{N}{2}]/H$\alpha$ map (Fig \ref{em_lines}). This region corresponds to a high velocity dispersion region for both the broad and narrow emission lines (Fig \ref{vdisp}), which suggest that the gas density in that region is higher and shock induced ionization dominates the emission lines. Since this region coincides with lower outflow velocities in its neutral phase, the higher density interstellar medium may be decelerating the outflow gas. 

\citet{rupke05b} showed that ULIRG winds have wider opening angles ($\sim3-4\pi$~steradians) than lower luminosity galaxies based on the detection rate of winds in ULIRGs, which is almost twice as high as in less luminous galaxies. The inconsistency between our outflow velocity maps and a basic disk wind model supports the hypothesis that the morphology of ULIRG winds may be different from winds of disk galaxies. The asymmetry of the outflow velocity maps may be a result of disturbance in the ISM due to merging.  Thus, a merger may affect the dynamics of the wind, causing it to deviate from a simple disk wind structure. 

The above discussion assumes that the outflowing material comes from the starbursting disk. It is possible that the outflow may be dominated by gas entrained outside of the disk, which may have a different velocity pattern due to the complex merger dynamics. For instance, the edges of our systemic velocity map show deviations from simple rotation.  However, we show below that the outflowing gas mass is high, suggesting that an origin in the massive central gas reservoir is most plausible. Furthermore, the outflow is dusty, implying that there is neutral, perhaps even molecular, gas in the wind, most likely originating in the dusty molecular disk.

\subsection{An Extended Multi-Phase Outflow}

An outstanding issue in galactic wind research is the relationship of different phases of the wind to each other.  Ionized gas has sometimes been seen in outflow in concert with neutral gas in starburst-powered ULIRG winds \citep{veilleux95,heckman00,rupke05b}.  Globally, ULIRG winds are also thought to be dusty, due to the strong positive correlation between equivalent width of the Na I D line and E(B-V) in LIRGs and ULIRGs.

Using our data, we can convincingly demonstrate that the ionized gas is outflowing on kiloparsec scales, with high velocities (exceeding 600 km~s$^{-1}$ in some locations), large linewidths (up to 600 km~s$^{-1}$), and strong collisional excitation. \citet{spoon09} have detected blueshifted ionized gas in ULIRGs with AGNs using mid-infrared emission lines. We have shown that starburst ULIRGs can also possess ionized outflows. As discussed in \S\ 5.1, the correlation with the neutral gas is weak, however, suggesting that the relationship between the ionized and neutral gas is complex.  Observations of more resolved sources may be necessary to understand the physical link between the two.

Our data also show, for the first time, a strong spatial correlation between dust reddening and NaI D equivalent width, again on kiloparsec scales.  This robustly confirms the physical coincidence of the neutral gas and dust in these outflows.  By inference, it also suggests that the filamentary dust features seen in HST images of F10565+2448 and other ULIRGs are in many cases instances of outflowing dust and gas.

This result raises again the tantalizing possibility that starburst- and AGN-driven winds are responsible for removing the cocoon of dust around obscured AGN in mergers.  Mid-infrared spectra of ULIRGs reveal deeply-absorbed continua \citep{veilleux09}; these large optical depths have been hypothesized to hide nascent quasars \citep{sanders88}.  Outflows like the one observed in this source may be the mechanism by which this dust is removed and a quasar is born.

\subsection{Outflow Parameters}
We can refine the estimation of mass and kinetic energy in the neutral wind using a wind model that depends on the physical parameter outputs of our fitting. We choose a spherical thin shell instantaneous outflow similar to the one used by \citet{rupke05b}. The difference is that \citeauthor{rupke05b} assumed spherically symmetric velocity and density, but we have information on the spatial variations of these quantities. An alternative to the thin shell model is an outflow with finite thickness and velocity that depends on radius. In \citet{veilleux94} and \citet{martin09}, the thick shell outflow has $V \propto R^{n}$, where $R$ is the (spherical) radial distance from the nucleus and $V$ is the outflow velocity.  Because we observe the galaxy in a face-on orientation, we cannot easily probe variations in velocity with $R$, so we assume a single, characteristic value for $R$ but let $V$ vary across the outflowing shell according to the velocity output of NAFIT.

We have de-projected the velocities from observed to purely radial velocities. The instantaneous mass outflow rate, average outflow rate and total mass are:
\begin{equation}
dM/dt_{thin}^{inst} = \mu m_{p} R^{2} \sum \frac{N(\theta,\phi)}{dr} v(\theta,\phi) sin(\theta) d\theta d\phi
\end{equation}

\begin{equation}
dM/dt_{thin}^{avg} = \mu m_{p} R \sum N(\theta,\phi) v(\theta,\phi) sin(\theta) d\theta d\phi
\end{equation}

\begin{equation}
M_{thin} = \mu m_{p} R^{2} \sum N(\theta,\phi) sin(\theta) d\theta d\phi
\end{equation}

where $R$ is the radius of the thin shell, theta and phi are the usual spherical angular coordinates, $dr$ is the thickness of the shell ($dr << R$), $\mu$ is the mean molecular weight, $m_{p}$ is the mass of a proton, and $N$ is the column density of hydrogen.  For all our calculations we choose $R = 5$ kpc to be consistent with \citet{rupke05b}, and $dr = 0.5~\mathrm{kpc} << R$. Using the mass outflow rate we can calculate the energy and momenta outflow rates (see Rupke et al. 2005b), the values are summarized in Table \ref{outflowrates}. The column density of hydrogen, N(H), can be estimated from N(NaI) given a metallicity and ionization fraction:
\begin{equation}
N(H) = N(NaI)(1 - y)^{-1}10^{-(a+b)}
\end{equation}
where $y = (1 - NaI/Na) = 0.9$, $a = log[N(Na)/N(H)] = -5.69$ which we set to solar value, and $b = log[N(Na)/N(H)_{total}] - log[N(Na)/N(H)_{gas}] = -0.95$ is the assumed depletion. 

The values that we calculated for mass, energy, momentum and their outflow rates are within a factor of $3-4$ of the values in \citet{rupke05b}. The new mass, energy and momentum are smaller than what is predicted from long slit data, while the outflow rates are larger. With the new spatial information on velocity and column densities, the new values should be more accurate. The thin shell geometry that we assumed is, however, uncertain. Theory and simulation predicts a thin shell of warm gas behind the forward shock of the wind \citep[e.g.][]{castor75, suchkov94, strickland00}, but it is unclear whether it is accompanied by neutral gas. Since we have limited information on the radial depth of the wind, we cannot rule out a thick outflowing shell in which neutral clouds are continuously entrained in the wind. Switching from a thin shell to thick shell model will decrease our measured mass. Furthermore we are only considering the part of the wind where we have reliable spectral fitting; the actual wind probably extends further than what we observed and carries more mass.  

In some optically thick cases, our fitting routine finds only a lower limit for $\tau$ for one of the velocity components, which is set to 5 in our fits. This is applicable to $\sim 9$\% of the spectra used for this section's calculations, and may cause us to underestimate the mass. We check the extend of this effect by comparing our values to those calculated by using only one velocity component fit throughout. When only one component is used for fitting, only $< 2$\% of the fitted $\tau$ used for calculation are at the lower limit. The estimated mass and energy are slightly smaller than, but within a factor of 2 of the values presented in table \ref{outflowrates}. From the present data alone, we cannot rule out the possibility that optically thick lines cause a significant underestimate of the mass in the wind. However, we believe it is more likely that our mass estimates are not biased in this way because: (1) the saturated components in the current data arise primarily at low signal-to-noise; (2) the one-component fits, which have more sensitivity to the overall shape of the lines in the low S/N region did not give a drastically different mass estimate. Higher S/N data will likely resolve this issue and better constrain the optical depths of each component. 

Since our observation only trace the NaI D clouds, it is possible that we underestimated the total mass of the wind by not considering the other phases. \citet{rupke05b} compared ratio of cold, neutral gas outflow rates to the expected mass injection rated from the hot gas created by supernovae. The ratios are found to be close to unity on average but it can vary from 0.001 to 10 for individual galaxies. On the other hand, theory suggests that the entrained clouds dominate the mass budget of the wind \citep[e.g.][]{strickland00}, and we suspect that the NaI D clouds are the entrained gas \citep{rupke05b}. The conversion factor used to convert the neutral Na column density to that of the total Na has also been shown to be uncertain \citep{murray07}. However, the total mass in the cold gas cannot be more than an order of magnitude larger than our calculated value, otherwise the mass ratio between cold and hot gas will be too large \citep{strickland00, strickland10}. 

Within the above constraints, the present data put on firmer footing the coarse mass outflow rates measured by \citet{rupke05b}.  It is clear that these outflows are massive and energetic.  The neutral outflow rate in this particular system is comparable to the star formation rate.  However, this particular system has a higher outflow rate than the average value in ULIRGs \citep{rupke05b}.  Further spatially resolved and multiphase observations are necessary to understand the exact relationship between mass outflow rate and star formation rate in galactic superwinds.

\begin{table}[h]
\centering
Table \ref{outflowrates}\\
Outflow Characteristics
\begin{tabular}{c c c c}
\hline\hline
$ $ & Total  & Outflow~rate~(Instantaneous) &  Outflow~rate~(Average)  \\
\hline
Mass & $9.5$ & $3.1$ & $2.1$\\
Energy & $57.1$ & $43.3$  & $42.3$ \\
Momentum & $49.8$ & $36.0$  & $35.0$\\
\hline
\hline
\end{tabular}
\caption{Mass, energy, momentum and the outflow rates in logarithmic values. Column (1) The log of total quantity of mass, energy and momentum. Column (2) The log of instantaneous outflow rate. Column (3) The log of average outflow rate. The units are the following: mass - $M_\odot$, mass outflow rates - $M_\odot$ yr$^{-1}$, energy - erg, energy outflow rates - erg s$^{-1}$, momentum - dyne s, momentum outflow rates - dyne.}
\label{outflowrates}
\end{table}

\section{Conclusion}

We have presented Gemini integral-field data of a galactic wind in the starburst ULIRG F10565+2448 to study the structure of the wind in detail.We fitted the Na I D absorption lines and the H$\alpha$ and [\ion{N}{2}] emission lines across the $7\times6$ kpc field of view using multiple velocity components. The systemic rotation velocity determined from the narrow emission lines is consistent with the molecular gas disk rotation. The absorption-line gas velocities shows asymmetric outflows with velocity gradients along the major axis opposite to that of the systemic rotation. We see widespread evidence of shock ionization through collisional excitation of the gas and large velocity dispersions, with particularly strong shocks observed in the southeast corresponding to the region with lower absorption-line outflow velocities. This supports the notion of gas deceleration in the east, which should be accompanied by shocking. Our results strengthen the hypothesis that ULIRG winds may differ in structure from those in the more quiescent galaxies. 

We observed both neutral and ionized outflowing gas. The correlation between the outflow velocities of these two phases is weak, suggesting a complex relationship between the two. We found a spatial correlation between dusty region and high Na I D equivalent width region, confirming spatial coincidence of dust and neutral entrained outflow material. The presence of dust in the outflow raises the possibility that winds may be responsible for removing dust around obscured AGN and revealing a bright quasar.  

We calculated the mass, energy and momentum outflow rate assuming a spherical thin-shell outflow. The numbers that we obtained are in general agreement with those in \citet{rupke05a}. However, by studying the structure of winds in other starburst mergers, we should be able to improve on the current results.  First, by studying edge-on disks we will be able to better constrain the opening angles of these winds.  Second, by studying a sample of winds, we will place better constraints on the relationship between the ionized and neutral phases of the wind, and whether asymmetric velocity structures are common.  Finally, and perhaps most importantly, we must compare observations to detailed simulations of mergers that include realistic wind prescriptions -- such simulations do not yet exist, but are necessary for understanding detailed data sets like the present one.

The authors thank Jabran Zahid for sharing routines that were the framework for our stellar and emission-line fitting software. This paper used data from the NASA/ESA Hubble Space Telescope, and obtained this data from the NASA/ESA Hubble Space Telescope, and obtained fro m the Hubble Legacy Archive, which is a collaboration between the Space Telescope Science Institute (STScI/NASA), the Space Telescope European Coordinating Facility (ST-ECF/ESA) and the Canadian Astronomy Data Centre (CADC/NRC/CSA). This paper used data from the Sloan Digital Sky Survey (SDSS). Funding for the Sloan Digital Sky Survey (SDSS) has been provided by the Alfred P. Sloan Foundation, the Participating Institutions, the National Aeronautics and Space Administration, the National Science Foundation, the U.S. Department of Energy, the Japanese Monbukagakusho, and the Max Planck Society. DSNR is supported by a University of Hawaii start-up grant to Lisa Kewley, and by NSF CAREER grant AST07-48559.  This work is based on observations obtained at the Gemini Observatory, which is operated by the Association of Universities for Research in Astronomy, Inc., under a cooperative agreement with the NSF on behalf of the Gemini partnership.

 \end{document}